\documentclass{emulateapj}

\usepackage{graphicx}
\usepackage{psfig}

\slugcomment{To appear in the ApJSupp Spitzer special issue}

\shorttitle{Obscured and unobscured AGN in the {\em Spitzer} FLS}
\shortauthors{Lacy et al.}

\begin{document}


\title{Obscured  and unobscured active galactic nuclei
in the Spitzer Space Telescope First Look Survey}


\author{M.\ Lacy\altaffilmark{1}, 
L.J.\ Storrie-Lombardi \altaffilmark{1}, A.\ Sajina\altaffilmark{2},
P.N.\ Appleton\altaffilmark{1}, L.\ Armus\altaffilmark{1}, 
S.C.\ Chapman\altaffilmark{1},
P.I.\ Choi\altaffilmark{1}, 
D.\ Fadda\altaffilmark{1},
F.\ Fang\altaffilmark{1},
D.T.\ Frayer\altaffilmark{1}, I.\ Heinrichsen\altaffilmark{1}, 
G.\ Helou\altaffilmark{1}, M.\ Im\altaffilmark{3}, 
F.R.\ Marleau\altaffilmark{1},
F.\ Masci\altaffilmark{1}, D.L.\ Shupe\altaffilmark{1}, 
B.T.\ Soifer\altaffilmark{1}, 
J.\ Surace\altaffilmark{1}, H.I.\ Teplitz\altaffilmark{1}, 
G.\ Wilson\altaffilmark{1}, L.\ Yan\altaffilmark{1}} 
\altaffiltext{1}{Spitzer Science Center, Caltech, MS 220-6,
Pasadena, CA 91125; mlacy,lisa@ipac.caltech.edu}
\altaffiltext{2}{Department of Physics \& Astronomy, University of British 
Columbia, 6224 Agricultural Road, Vancouver, BC V6T1Z1, Canada, 
sajina@astro.ubc.ca }
\altaffiltext{3}{Astronomy Program, School of Earth and Environmental 
Sciences, Seoul National University, Shillim-dong, Kwanak-gu, Seoul, S. Korea
2-880-9010}

\begin{abstract}
Selection of active galactic nuclei (AGN) in the infrared allows the discovery
of AGN whose optical emission is extinguished by dust. In this paper, 
we use the {\em Spitzer Space Telescope} First Look Survey (FLS) to assess 
what fraction of AGN with mid-infrared luminosities comparable to quasars are 
missed in optical quasar surveys due to dust obscuration. We begin by 
using the Sloan 
Digital Sky Survey (SDSS) database to identify 54 quasars within the  
4 deg$^{2}$ extragalactic FLS. These quasars occupy a distinct 
region in mid-infrared color space by virtue of their strong, red, continua. 
This has allowed us to define a mid-infrared color criterion for selecting 
AGN candidates. About 2000 FLS objects have colors consistent with them 
being AGN, but most are much fainter in the mid-infrared than the SDSS 
quasars, which typically have 8$\mu$m flux densities, $S_{8.0}$, $\sim 1$mJy. 
We have investigated the properties of the 43 objects with $S_{8.0} \geq 1$mJy
satisfying our AGN color selection. This sample should contain both 
unobscured quasars, and AGN which are absent from the SDSS survey due to 
extinction in the optical. After removing 16 known quasars, three probable
normal quasars, and eight spurious or confused objects from the 
initial sample of 43, we are left with 16 objects which are likely to be 
obscured quasars or luminous Seyfert-2 galaxies. This suggests the numbers 
of obscured and unobscured AGN are similar in samples selected in the 
mid-infrared at $S_{8.0}\sim 1$mJy.

\end{abstract}



\section{Introduction}

The number
of AGN missed in optical or soft X-ray surveys due to obscuring 
columns of dust and gas in the host is still an open question (e.g., 
Webster et al.\ 1995; Cutri et al.\ 2001; Gregg
et al.\ 2003; Norman et al.\ 2002; Richards et al.\ 2003). 
Understanding and quantifying the population of obsured AGN is
important if we are to obtain a 
complete picture of the build-up of black holes in the nuclei of 
galaxies. The close link between bulge mass and black hole mass (e.g.\
Magorrian et al.\ 1998) implies that understanding this history 
will also help us to understand the galaxy formation process as a whole.

The local mass density in black holes 
is dominated by the $\sim 10^{8.5}M_{\odot}$ black
holes in $L^{*}$ elliptical galaxies. When accreting at Eddington rates, 
these correspond to luminous quasars ($M_{B}\approx -25$). Recent
work on the X-ray background, however, indicates that most of the mass 
build-up may occur on longer timescales at sub-Eddington rates, mostly in 
obscured AGN (Cowie et al.\ 2003). Besides X-ray searches, other techniques
for finding obscured AGN have also been employed.
A sample of luminous narrow-line AGN has been selected from the SDSS 
(Zakamska et al.\ 2003). Quasar
samples selected in the near-infrared
from the Two Micron All-Sky Survey (2MASS) 
(Cutri et al.\ 2001; Glikman et al.\ 2004) 
have identified a number of quasars 
with significant optical extinction and X-ray columns. Even in 
the mid-infrared (MIR), a highly-obscured AGN will still have significant 
extinction. However, the disk or torus-like geometry of the obscuring material
means that even if the extinction to 
the broad-line region is high, MIR light will usually be visible, 
unless the geometry is truly edge-on, or 
the AGN is completely enveloped in a dense cloud of dust and gas
(Pier \& Krolik 1993; Efstathiou \& Rowan-Robinson 1995). 
Thus an MIR survey can usefully address the question
of the number of obscured AGN.

Studies with the {\em Infrared Space Observatory} (ISO) have shown that the
strong MIR continuum associated with AGN provide a unique spectral 
signature that can be used to distinguish AGN from starbursts.
The MIR continuum from galaxies 
arises mostly from three sources: H{\sc ii} regions dominated by emission 
from very small dust grains (producing a steeply-rising continuum at 
12-16$\mu$m), photodissociation regions dominated
by bands of Polycyclic Aromatic Hydrocarbon (PAH) emission, and 
AGN dominated by a strong 3-10$\mu$m continuum. 
Laurent et al.\ (2000) show that the spectral energy distributions (SEDs)
arising from each of these is sufficiently distinct to allow discrimination
of AGN from star-forming galaxies based on their MIR SEDs. Using these ideas,
Haas et al.\ (2004) combined near- and mid-infrared colors to 
select AGN from an {\em ISO} parallel survey.
With the advent of {\em Spitzer} (Werner et al.\ 2004) it has
become possible to obtain MIR photometry for large samples of 
field galaxies, and thus to use the MIR SEDs to identify specific 
populations of AGN and star-forming galaxies. 
In this paper we use SDSS quasars to provide an empirical localization of the 
AGN population in mid-infrared color space. We then select a sample of 
candidate obscured 
AGN with 8$\mu$m flux densities ($S_{8.0}$) $\geq 1$mJy, and compare their
optical identifications and estimated redshifts to those of 
the SDSS quasars with $S_{8.0}\geq 1$mJy. We use the
main field of the  extragalactic component
of the {\em Spitzer} First Look Survey (FLS), a shallow, 4 deg$^2$ 
survey with the Infrared Array Camera (IRAC; Fazio et al.\ 2004) 
and the Multiband Imaging
Photometer for {\em Spitzer} (MIPS; Rieke et al.\ 2004), for which 
an extensive multiwavelength ancillary dataset exists. 
We assume a cosmology with $\Omega_{\rm M}=0.3$, $\Omega_{\Lambda}=0.7$
and $H_0=70 {\rm kms^{-1}Mpc^{-1}}$.

\section{The Spitzer FLS main field dataset}

The FLS observations were made in December 2003 (program ID 26,
astronomical observation request (AOR) IDs 
\dataset[ads/sa.spitzer/#0003861504]{3861504}, 
\dataset[ads/sa.spitzer/#0003861760]{3861760}, 
\dataset[ads/sa.spitzer/#0003862016]{3862016},
\dataset[ads/sa.spitzer/#0003862016]{3862016},
\dataset[ads/sa.spitzer/#0003862272]{3862272}, 
\dataset[ads/sa.spitzer/#0003862528]{3862528}, 
\dataset[ads/sa.spitzer/#0003862784]{3862784},
\dataset[ads/sa.spitzer/#0003863040]{3863040}, 
\dataset[ads/sa.spitzer/#0003863296]{3863296},
\dataset[ads/sa.spitzer/#0003863552]{3863552}). 
We have used preliminary versions of 
the IRAC catalog (Lacy et al.\ 2004) and 
the MIPS 24$\mu$m catalog (Fadda et al.\ 2004b). The IRAC catalog 
has flux density limits ($5\sigma$ in a 5$^{''}$ diameter
aperture) in the four IRAC channels with nominal central wavelengths
of 3.6$\mu$m, 4.5$\mu$m, 5.8$\mu$m and 8.0$\mu$m, of $S_{3.6}\approx 7\mu$Jy, 
$S_{4.5}\approx 8\mu$Jy, $S_{5.8}\approx 60\mu$Jy, and 
$S_{8.0}\approx 50\mu$Jy. The MIPS 24$\mu$m catalog has a flux density 
limit of $\approx 300\mu$Jy. 

\section{The SDSS quasars}

The Sloan Data Release 1 (DR1) quasar survey (Schneider et al.\ 2003)
contains 54 quasars which fall within the main field of the
FLS. All are detected with IRAC, and all but one
at 24$\mu$m with MIPS. Three more quasars were found as a result
of spectroscopic follow-up of the radio survey of Condon et al.\ (2003) using 
the Kast Spectrograph at Lick Observatory.

\section{The position of SDSS quasars in MIR color-color plots}

An IRAC color-color plot, using all four 
broad-band channels of the IRAC instrument, is shown in Figure 1. The dots
indicate the location of $\approx 16000$ objects in the main field catalog.
Plotting the
8.0$\mu$m/4.5$\mu$m ratio against the 5.8$\mu$m/3.6$\mu$m ratio makes the 
color-color plot effective at separating objects with blue continua
from those with red. Most objects have blue colors in both axes. 
These are most likely stars, or low-redshift galaxies whose SEDs are weak in 
non-stellar light, e.g., elliptical galaxies. From this clump 
two distinct ``sequences'' can be seen. One has blue colors in 
$S_{5.8}/S_{3.6}$ and very red colors in
$S_{8.0}/S_{4.5}$. These are most likely low-redshift 
($z\stackrel{<}{_{\sim}} 0.2$) 
galaxies with the centers of their 6.2 and 7.7$\mu$m PAH emission bands
(Puget \& L\'{e}ger 1989) redshifted into the IRAC 
8.0$\mu$m filter. The other sequence has red colors in both pairs of 
filters and it 
is on this sequence that the SDSS and radio-selected quasars lie. Note 
that strong stellar light will shift the $S_{5.8}/S_{3.6}$ ratio to the 
blue, as illustrated by the two filled squares in Figure 1. 
These objects are classed as quasars in the SDSS database, but have
resolved host galaxies in the IRAC images, and so are classed as Seyfert-1s in
this study.

\begin{figure}

\plotone{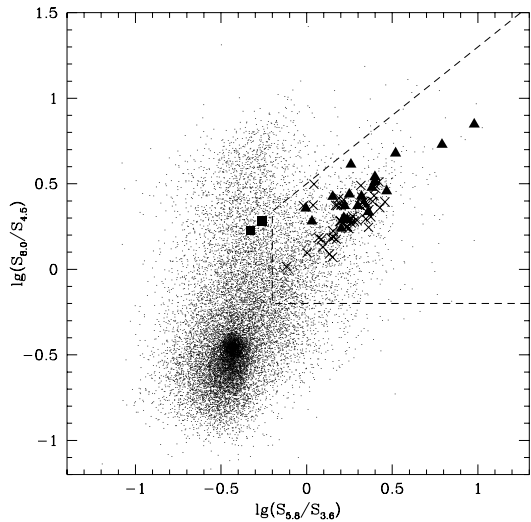}

\caption{An IRAC color-color plot using the main field FLS data. 
Dots represent the $\approx 16000$ objects with ``clean''
detections in all four IRAC bands. Crosses indicate the colors of all the
SDSS and radio-selected
quasars, squares the SDSS Seyfert 1 galaxies and triangles the 
bright ($S_{8\mu m}\geq
1$mJy) sample of obscured AGN. The dashed line shows the color criteria 
used to pick out the AGN sample.} 
\end{figure}

\begin{figure}

\plotone{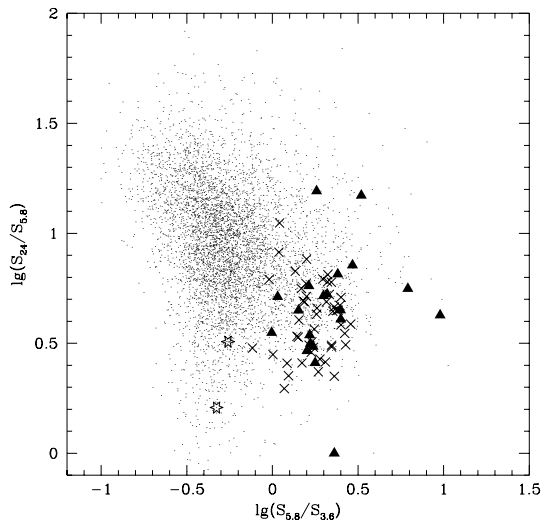}

\caption{An IRAC-MIPS 24$\mu$m color-color plot. Dots represent objects 
with ``clean'' detections in all four IRAC bands matched to the MIPS
24$\mu$m catalog of Fadda et al.\ (2004b), totalling $\approx 6000$ 
objects. Crosses indicate the colors of SDSS
quasars, stars the SDSS Seyfert 1 galaxies and triangles the $S_{8.0}\geq
1$mJy sample of obscured AGN candidates.} 
\end{figure}

\section{MIR selection of candidate AGN}

We used {\em Spitzer} colors of the SDSS quasars to empirically define a 
color selection which will pick out objects with MIR SEDs consistent with 
their being AGN. This is the region enclosed by the dashed line in Figure 1. 
Our modelling of MIR SEDs based on {\em ISO} spectra 
(Sajina, Lacy \& Scott 2004) suggests that most of the
objects in this region are indeed dominated by AGN emission, though we expect
some contamination by star-forming galaxies close to the boundaries.
There are $\approx 2000$ sources in this region, which are therefore
likely to contain AGN. Most of them are fainter in the MIR than the SDSS
quasars, which are typically much brighter than the FLS flux density limits. 

To pick a sample of AGN selected in the MIR which we could 
compare directly to the SDSS quasar sample, we examined the 
distribution of $S_{8.0}$ for the SDSS sample. (8.0$\mu$m is the longest 
IRAC wavelength, and therefore least affected by dust obscuration.)
This distribution showed a peak at 
$S_{8.0}\approx 1$mJy, and we therefore assumed that at $S_{8.0}\geq 1$mJy
most of our MIR-selected AGN would appear in the SDSS sample if they 
were not obscured by dust.

There are 43 objects within the AGN selection region of the color-color
plot with $S_{8.0} \geq 1$mJy and ``clean''
detections in the IRAC FLS catalogue [i.e.\ unblended (about 80\% 
of the 4-band detections), and outside the haloes of bright stars and 
unaffected by multiplexer bleed or pulldown 
(97.6\% of the survey area)]. 
The steps used to determine which of these were candidate obscured AGN
were as follows: (1) we matched the sample of 43 bright AGN candidates
to the SDSS quasar list, finding that 14 of our 43 objects are known SDSS 
quasars; (2) we next matched to our list of radio-selected quasars, finding
two objects in common; (3) we examined the objects by eye on the FLS mosaics,
and rejected a further seven objects from the sample on the basis of them
being either saturated stars, or confused in the IRAC images. 
This left 20 objects in the sample, which are our 
bright, obscured AGN candidates. These are shown as triangles in Figure 1,
and listed in Table 1. As a check on the AGN nature of these candidates, 
in Figure 2 we plot a color-color plot which extends
the wavelength range to MIPS 24$\mu$m. The large wavelength 
difference between 8 and 24$\mu$m makes the 
interpretation of the MIPS/IRAC color-color plots more complicated, but it 
can be seen that most of the obscured AGN candidates lie in the same
region of the plot as the SDSS quasars, consistent with them having 
SEDs dominated by emission from hot 
dust around the AGN. Starburst galaxies typically have 
much redder $S_{24}/S_{5.8}$ colors due to a lack of high temperature 
dust emission from the AGN.

All our 20 bright, obscured AGN candidates (and all the SDSS quasars)
were identified on the 
$R$-band images of Fadda et al.\ (2004a) (limiting  $R\approx 25.5$). 
These identifications allowed us to further refine our selection.  
Out of the 20 obscured AGN candidates, we found that three had the 
MIR to optical color and stellar morphology of 
normal quasars (one apparently behind a spiral disk), 
one candidate is very bright ($R=15.7$), 
so is probably either a peculiar star, or a 
foreground star within $\sim 1^{''}$ of a background 
AGN. Of the remaining 16 objects, 14 are extended in $R$-band, 
and thus most likely 
galaxies whose AGN is completely obscured in the optical. Two are faint in 
$R$, but point-like and so are probably only partly obscured AGN. Figure 3 
shows 
a plot of the ratio of $R$-band flux density ($S_{0.65}$) to $S_{8.0}$ 
plotted against lg($S_{8.0}$) for all 20 objects. 

\begin{table}
\caption{Candidate AGN not in SDSS DR1 with $S_{8.0}\geq 1$mJy}
\begin{tabular}{lccccl}
Name &$S_{8.0}^{1}$&$R$ &$S_{1.4{\rm GHz}}^{2}$&$z_{\rm phot}^{3}$&Opt \\
     &(mJy)       &     &(mJy)&                  &Type$^{4}$\\\hline
SSTXFLS J171106.8+590436&1.28&19.8&0.31& 0.460          & galaxy\\
SSTXFLS J171115.2+594906&4.53&21.0&0.12&  -             & stellar\\
SSTXFLS J171133.4+584055&1.91&18.6&0.09& - & see note (5)\\
SSTXFLS J171147.4+585839&1.56&21.0&0.57& (0.5)      & galaxy\\
SSTXFLS J171313.9+603146&4.30&18.3&$<$0.09  & 0.155      & galaxy\\
SSTXFLS J171324.3+585549&1.09&20.7&$<$0.09  & 0.635      & galaxy\\
SSTXFLS J171325.1+590531&1.19&18.3&0.16& 0.105          & galaxy\\
SSTXFLS J171421.3+602239&1.25&18.6&0.16& 0.195          & galaxy\\
SSTXFLS J171430.7+584225&1.85&19.9&0.16& 0.135          & galaxy\\
SSTXFLS J171708.6+591341&1.32&21.1&$<$0.09  & 0.195          & galaxy\\
SSTXFLS J171804.6+602705&1.05&21.7&$<$0.09  & (0.6)          & galaxy\\
SSTXFLS J171831.6+595317&1.10&20.6&2.22& 0.295          & galaxy\\
SSTXFLS J171930.9+594751&1.44&19.6&0.56&0.355          & galaxy\\
SSTXFLS J172050.4+591511&3.35&21.6&1.75& 0.440          & galaxy\\
SSTXFLS J172123.1+601214&3.40&18.8&0.26& 0.355          & galaxy\\
SSTXFLS J172253.9+582955&1.00&18.5&$<$0.09  &- & stellar\\
SSTXFLS J172328.4+592947&1.48&22.2&0.31 & -             & stellar\\
SSTXFLS J172432.8+592646&3.04&15.7&$<$0.09 & -             & stellar$^{6}$\\
SSTXFLS J172458.3+591545&1.10&20.0&0.46&0.640          & galaxy\\
SSTXFLS J172601.8+601100&1.43&20.3&$<$0.09 & -             & stellar\\
\end{tabular}

\noindent
Notes: (1) preliminary flux densities only, errors $\approx 20\%$; (2)
flux densities at 1.4GHz in the survey of Condon et al.\ (2003);
(3) photometric redshifts, $z_{\rm phot}$, those 
in brackets are obtained by assuming a galaxy 
luminosity of $L^{*}$, others are from Hyperz using the SDSS $ugri$
magnitudes; (4) optical classification in the $R$-band image; (5) this
is a stellar object within the isophotes of a spiral disk, we assume
it is a background quasar which happens to be behind a disk galaxy; (6)
this object may be a star, see text. 
\end{table}

\begin{figure}

\plotone{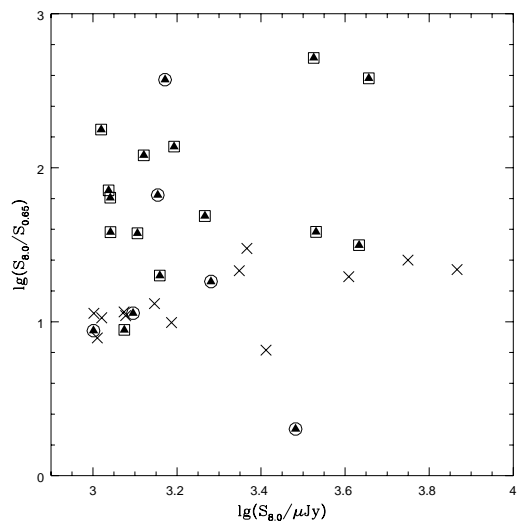}

\caption{$S_{8.0}$ to $R$-band flux ratio as a function of $S_{8.0}$.
SDSS quasars are shown as crosses, point-like identifications of AGN 
candidates are shown as triangles within circles, and triangles
within squares denote galaxy identifications.}
\end{figure}

\section{Photometric redshifts of the candidate AGN population}

Of the 14 extended (galaxy) detections in $R$-band, 
12 are detected on the SDSS DR1 images (which are significantly deeper than 
the SDSS spectroscopic quasar and galaxy surveys), and we have used their 
{\em ugri} magnitudes to make a rough photometric redshift estimates
using {\em Hyperz} (Bolzonella, Miralles \& Pell\'{o} 2000), with the
Bruzual \& Charlot (1993) templates. 
(We exclude the $z$-band due to the 
likelihood of contamination by AGN emission, but believe the other 
bands should be largely free of significant contamination given the extended
appearance of the galaxies in the $R$-band data). The galaxies have a mean 
luminosity of $\approx L^{*}$, and inspection of the $R$-band image shows 
they appear to be mostly early-type galaxies. Our photometric
redshifts are only approximate, but the magnitudes of the galaxies are 
such that few are likely to be at $z\stackrel{>}{_{\sim}}0.7$. 

\section{Discussion}

Although we emphasize that the candidate obscured AGN we have identified
from their MIR colors require spectroscopic confirmation as AGN, our 
study suggests that about 50\% of AGN selected in the MIR at these flux 
levels may be sufficiently obscured by dust in the optical for them
to be missed from the SDSS quasar survey. This excludes
AGN which are not energetically dominant in their host galaxies 
(which would plot elsewhere on the IRAC
color-color diagram), and any AGN which remain obscured in the MIR. 

The median photometric redshift of our candidate obscured AGN is 
$\approx 0.36$ (assuming the faint identifications with stellar 
morphologies to lie at high redshift), 
compared to $\approx 0.69$ for the SDSS quasars with $S_{8.0}\geq 1$mJy. 
There is thus a hint that the obscured AGN may have lower redshifts and 
luminosities. 
Selection effects are unlikely to account for this difference.
MIR selection ensures that the AGN are selected independent of their
optical properties. Whilst we might be biased towards objects with 
more dust close to the AGN, this would not introduce a redshift bias if the
fraction of such objects was constant with both luminosity and redshift.
A lower median luminosity for the obscured AGN would, however, 
be consistent with the 
``receding torus'' model (Lawrence 1991) in which a higher fraction of 
low-luminosity AGN are obscured, 
and might also be consistent with some studies
of the X-ray background which find large numbers of low luminosity, relatively
low redshift obscured AGN 
(Gilli et al.\ 2001; Gandhi \& Fabian 2003; Steffen et al.\ 2003). 

If unobscured, our AGN would be moderately
luminous, with $B$-band absolute magnitudes, $M_{B}$, $\approx -22$, on the 
boundary between Seyfert-1
galaxies and quasars. The brightest, $z\sim 0.6$, objects would have
$M_{B}\sim -23$, in the quasar regime, 
and close to the median luminosity of the SDSS sample. Pursuing 
this work to lower 8$\mu$m flux densities, and using spectroscopy
to confirm the AGN nature and redshift distribution of our objects,
will help us link the infrared to the X-ray studies. This will allow us to 
derive a consistent story for the nature and 
evolution of obscured AGN, and thus for how and when the supermassive 
black holes in galaxies today accreted most of their mass.

\acknowledgments

We thank Michael Gregg for assistance with 
the Lick observations, Susan Ridgway for advice and comments, 
B.\ Januzzi and A.\ Ford for 
the $R$-band survey, and J.\ Condon for the VLA 
survey. We thank the SDSS team for targeting the FLS region in the DR1.
This work is based on observations made with the 
{\em Spitzer Space Telescope},
operated by the Jet Propulsion Laboratory (JPL), California Institute
of Technology under NASA contract 1407. Support was provided
by NASA through JPL. The SDSS 
Archive is funded by the Alfred P. Sloan Foundation, the 
Participating Institutions, NASA, 
the National Science Foundation, the U.S. Department of 
Energy, the Japanese Monbukagakusho, and the Max Planck Society.









\begin{thebibliography}{}
\bibitem[]{} Bolzonella, M., Miralles, J.-M.\ \& Pell\'{o}, R.\ 2000, A\&A, 363, 476
\bibitem[]{} Bruzual, G.\ \& Charlot, S.\ 1993, ApJ 405, 538 
\bibitem[]{} Condon, J.J., Cotton, W.D., Yin, Q.F., Shupe, D.L.,
Storrie-Lombardi, L.J., Helou, G., Soifer, B.T.\ \& Werner, M.W.\ 
2003, AJ, 125, 2411 
\bibitem[]{} Cowie, L.L., Barger, A.J., Bautz, M.W., Brandt, W.N.\ \& Garmire,
G.P.\ 2003, ApJ, 584, L57
\bibitem[]{} Cutri, R.M., Nelson, B.O., Kirkpatrick, J.D., Huchra, J.P.\ \& 
Smith, P.S.\ 2001, in ASP Conf. Ser. 232, The New Era of Wide Field Astronomy,
ed. R.\ Clowes, A.\ Adamson, \& G.\ Bromage (San Francisco: ASP), 78
\bibitem[]{} Efstathiou, A.\ \& Rowan-Robinson, M.\ 1995, MNRAS, 273, 649
\bibitem[]{} Fadda, D., Jannuzi, B., Ford, A.\ \& Storrie-Lombardi, L.J.\ 
2004a, AJ, in press (astro-ph/0403490)
\bibitem[]{} Fadda, D.\ et al.\ 2004b, in preparation
\bibitem[]{} Fazio, G. G. et al. 2004, ApJS, this volume.
\bibitem[]{} Gandhi, P.\ \& Fabian, A.C.\ 2003, MNRAS, 339, 1095
\bibitem[]{} Gilli, R., Salvati, M.\ \& Hasinger, G.\ 2001, A\&A 366, 407
\bibitem[]{} Glikman, E., Gregg, M.D., Lacy, M., Helfand, D.J., Becker, R.H.\
\& White, R.L.\ 2004, ApJ, in press (astro-ph/0402386) 
\bibitem[]{} Gregg, M.D., Lacy, M., White, R.L., Glikman, E., Helfand, D., 
Becker, R.H. \& Brotherton, M.S.\ 2002, ApJ, 563, 133
\bibitem[]{} Haas, M., Siebenmorgen, R., Leipski, C., Meusinger, H., 
M\"{u}ller, S.A.H., Chini, R.\ \& Schartel, N.\ 2004, A\&A, in press (astro-ph/0404306) 
\bibitem[]{} Lacy, M.\ et al.\ 2004, in preparation
\bibitem[]{} Laurent, O., Mirabel, I.F., Charmandaris, V., Gallais, P., 
Madden, S.C., Sauvage, M., Vigroux, L.\ \& Cesarsky, C.\ 2000, A\&A, 359, 887
\bibitem[]{} Lawrence, A.\ 1991, MNRAS, 252, 586
\bibitem[]{} Magorrian, J.\ et al.\ 1998, AJ, 115, 2285
\bibitem[]{} Norman, C.\ 2002, ApJ, 571, 218
\bibitem[]{} Pier, E.A.\ \& Krolik, J.H.\ 1993, ApJ, 418, 673
\bibitem[]{} Puget, J.-L.\ \& L\'{e}ger, A.\ 1989, Ann. Rev. A. Ap. 27, 161
\bibitem[]{} Richards, G.T.\ et al.\ 2003, AJ, 126, 1131
\bibitem[]{} Rieke, G. H. et al. 2004, ApJS, this volume.
\bibitem[]{} Sajina, A., Lacy, M.\ \& Scott, D.\ 2004, in preparation
\bibitem[]{} Schneider, D.P.\ et al.\ 2003, AJ, 126, 2579
\bibitem[]{} Steffen, A.T., Barger, A.J., Cowie, L.L., Mushotzky, R.F.\ 
\& Yang, T.\ 2003, ApJ, 596, L23
\bibitem[]{} Webster, R.L., Francis, P.J., Peterson, B.A., Drinkwater, M.J., 
Masci, F.J.\ 1995, Nature, 375, 469
\bibitem[]{} Werner, M.\ et al.\ 2004, ApJS, this volume
\bibitem[]{} Zakamska, N.L.\ et al.\ 2003, AJ, 126, 2125

\end{thebibliography}
\end{document}